\documentclass[12pt,preprint]{aastex}

\shorttitle{Variable stars in the Hercules dSph} 
\shortauthors{Musella et al.}

\begin{document}


\title{Stellar Archaeology in the Galactic halo with Ultra-Faint Dwarfs: VII.  Hercules\altaffilmark{1}}


\author{Ilaria Musella\altaffilmark{2},  
Vincenzo Ripepi\altaffilmark{2},  
Marcella Marconi\altaffilmark{2}, 
Gisella Clementini\altaffilmark{3}, 
Massimo Dall'Ora\altaffilmark{2}, 
Victoria Scowcroft\altaffilmark{4,5},
Maria Ida Moretti\altaffilmark{6},
Luca Di Fabrizio\altaffilmark{7}, 
Claudia Greco\altaffilmark{8}, 
Giuseppina Coppola\altaffilmark{2},
David Bersier\altaffilmark{9},
M\'arcio Catelan\altaffilmark{10}, 
Aniello Grado\altaffilmark{2}, 
Luca Limatola\altaffilmark{2}, 
Horace A.Smith\altaffilmark{11},  
\and
Karen Kinemuchi\altaffilmark{12}}

\altaffiltext{1}{Based on data collected at the 2.5 m Isaac Newton Telescope, La Palma, Canary Island, Spain, at the 4.2 m William 
Herschel Telescope, Roche de los Muchachos, Canary Islands, Spain, at the 2.2 m ESO/MPI telescope, La Silla, Chile, Proposal 079.D-0587,
at the 2 m Liverpool Telescope, Roche de los Muchachos, Canary Islands, Spain, and at the 2 m Faulkes Telescope North,  Haleakala Observatory, Hawaii, USA}
 \altaffiltext{2}{INAF, Osservatorio Astronomico di Capodimonte,
   Napoli, Italy, ilaria@na.astro.it, ripepi@na.astro.it,
   marcella@na.astro.it,  dallora@na.astro.it, coppola@na.astro.it} 
 \altaffiltext{3}{INAF, Osservatorio Astronomico di Bologna, Bologna, Italy; gisella.clementini@oabo.inaf.it}
 \altaffiltext{4}{Observatories of the Carnegie Institution of Washington 813 Santa Barbara St., Pasadena, CA 91101}
 \altaffiltext{5}{Astrophysics Research Institute, Liverpool John
   Moores University, Birkenhead, UK; vs@obs.carnegiescience.edu} 
\altaffiltext{6}{Dipartimento di Astronomia, Universit\'a di Bologna, Bologna, Italy; mariaida.moretti@studio.unibo.it}
 \altaffiltext{7}{INAF, Centro Galileo Galilei \& Telescopio Nazionale Galileo, S. Cruz de La Palma, Spain; difabrizio@tng.iac.es} 
 \altaffiltext{8}{Observatoire de Gen\`eve, 51, ch. Des Maillettes, CH-1290 Sauverny, Switzerland; claudia.greco@obs.unige.ch} 
 \altaffiltext{9}{Astrophysics Research Institute, Liverpool John Moores University, Birkenhead, UK}
 \altaffiltext{10}{Pontificia Universidad Cat$\rm{\acute{o}}$lica de
   Chile, Facultad de Fisica, Departamento de Astronom\'{\i}a y Astrof\'{\i}sica,  Santiago, Chile; mcatelan@astro.puc.cl}
\altaffiltext{11}{Department of Physics and Astronomy, Michigan State University, East Lansing, MI 48824, USA; smith@pa.msu.edu}
 \altaffiltext{12}{NASA-Ames Research Center/Bay Area Environmental Research Institute Mail Stop 244-30, P.O. Box 1, Moffett Field, CA 94035-0001; karen.kinemuchi@nasa.gov} 

\begin{abstract}

We  present the first time-series study   of the ultra-faint dwarf
galaxy Hercules. Using a variety of telescope/instrument facilities we
secured about 50 $V$ and 80 $B$ epochs. These data allowed us to detect and
characterize 10 pulsating variable stars in Hercules.  Our final sample
includes 6 fundamental-mode (ab-type) and 3 first overtone (c-type) RR
Lyrae stars,  and one Anomalous Cepheid.
The average period of the ab-type RR Lyrae stars,
$\langle$P$_{ab}\rangle $=0.68 d ($\sigma$ = 0.03 d), places
Hercules in   the Oosterhoff II group, as found for almost the
totality  of the ultra-faint dwarf galaxies investigated so far for
variability. The RR Lyrae stars were used to obtain independent estimates
of the metallicity, reddening and distance to Hercules, for which we
find: $[Fe/H]=-2.30\pm0.15$ dex, $E(B-V)=0.09\pm0.02$ mag, and
$(m-M)_0=20.6 \pm 0.1$ mag, in good agreement with
the literature values.
We have obtained a $V, B-V$ color-magnitude diagram (CMD) of
Hercules that reaches  $V \sim$ 25 mag and  extends beyond the
galaxy's half-light radius over a total area of $40' \times 36'$. 
 The CMD and the RR Lyrae stars indicate the presence of
a population  as old and metal-poor as (at least) the Galactic
globular clusters M68. 

\end{abstract}

\keywords{galaxies: dwarf
---galaxies: individual (Hercules)
---stars: distances
---stars: variables: other
---techniques: photometric}

\section{Introduction}

The discovery of a new class of faint   dwarf satellites   around  the
Milky  Way (MW; see e.g., \citealt[and references
therein]{bel06,Bel10})  and the  Andromeda galaxies (M31; see e.g.,
\citealt[and references therein]{Richardson11})   has opened a new
window for the study of the formation history of   large spirals.  

The new systems show a number of remarkable differences with respect
to  the ``classical'' dwarf spheroidals (dSphs) surrounding the MW and
M31: i) they  have much  lower surface brightnesses  ($\mu_V \gtrsim
28$~mag),  for which they were named ``ultra-faint'' dwarfs  (UFDs);
ii) they are very metal poor, with large dispersions and [Fe/H] values
as low as $-4$ dex \citep[see][and references therein]{THT09}. Such
extreme abundances are not observed   among the classical dSphs where
only a few stars with [Fe/H]$<-3.0$ have  been detected  \citep{Fr10}
compared  to the large number found in the MW halo; iii)  they
generally contain RR Lyrae stars that conform to the subdivision into
Oosterhoff types I (Oo I) and II \citep[OoII;][]{Oo39}\footnote{The
Galactic globular clusters can be divided into two different groups
according to the mean period of their  ab-type  RR Lyrae stars:
Oosterhoff type I (Oo I) clusters have $\langle P \rangle=0.55$ d,
whereas type II  (Oo II) clusters have  $\langle P \rangle=0.65$ d}
observed for field and cluster MW variables.  So far, the only
exception among the UFDs is Canes Venatici I \citep[CVn
I,][]{Kuehn08}, the brightest of the MW UFDs, that, like the classical
MW dSphs,  has instead Oosterhoff-intermediate (Oo-Int) properties
\citep[][and references therein]{Cat09,Clementini10};
and, finally: iv) the UFDs discovered so
far outnumber by almost a factor of two the classical dSphs, thus
partially reducing the so-called ``missing satellites problem''
\citep{Moore99,Kl99} affecting the $\Lambda$-cold-dark-matter
($\Lambda$CDM) scenario of galaxy formation.  With their properties
the UFDs are potentially much better analogues than the classical dSphs of  the
``building blocks''  that contributed to the formation of the two
large spirals in the Local Group.  They have absolute luminosities
similar to the globular clusters (GCs; $M_V\sim -7$ mag, on average) but
they are much more spatially extended than GCs.  With typical half-light radii of
$r_h \ge$ 100 pc, in fact they equal in size the ``classical'' dSphs.
The UFDs are found in groups on the sky  \citep[see e.g. Fig. 1
of][]{Richardson11},  have small velocity
dispersions, and  high mass-to-light ratios.    

All UFDs host an ancient population around 10 Gyr old. They have
GC-like color-magnitude diagrams (CMDs) resembling  the CMDs of
metal-poor Galactic GCs (GGCs) such as M92 (NGC~6341), M15 (NGC~7078)
and M68 (NGC~4590).  Some
of the MW UFDs have a  distorted shape due to the tidal interaction
with the MW.

We are carrying out an extensive observational campaign  of  the new
MW and M31 UFDs  to study structural parameters and stellar
population properties, as well as the variable stars of these
systems. We have already published results  for Bootes I \citep[Boo
I,][]{Dallora06}, CVn I \citep[][]{Kuehn08}, Canes
Venatici II \citep[CVnII,][]{Greco08}, Coma \citep{Musella09}, Leo IV
\citep{Moretti09},  and Ursa Major II \citep[UMaII][]{Dallora12} among the
MW UFDs.  With the only exception of CVn I, all these dwarfs contain RR Lyrae stars  with pulsation
periods suggesting an Oo II classification. 
However, only Boo I and CVn I contain
sufficiently large numbers of variables to be safely classified into
Oosterhoff types. The  classification of the other UFDs is less
certain given the small numbers of  variables they
contain. Nevertheless,  their few  variables clearly tend to have
Oosterhoff type II properties. Thus, in terms of stellar metallicity
and pulsation properties of the variable stars, systems similar to the
UFDs, as they were at earlier times, could resemble the building
blocks of the Galactic halo.

In this paper we extend our analysis to the Hercules UFD galaxy
(R.A.$=16^h 31^m 02.0^s$, Decl.$=12^{\circ}47^{\prime}29.6^{"}$,
J2000.0).   The galaxy was discovered by \citet{Bel07} from the
analysis of SDSS data.  The CMD of Hercules (based on follow-up Isaac
Newton Telescope - INT- data) shows besides the red giant branch
(RGB), also blue and red horizontal branches (HBs).  \citet{Bel07}
interpreted this evidence as a possible signature of multiple stellar
populations present in the galaxy.  A number of  studies  have been
devoted to this galaxy after the discovery paper.  CMDs of
Hercules reaching well below the galaxy's main sequence turn-off  were
published by \citet[][in the $B$ and $V$ bands]{Col07} and \citet[][in
the $g$ and $B$ bands]{Sand09}, based on very  wide-field
observations obtained with the  red and blue channels of the Large
Binocular Camera (LBC; \citealt{Giallongo08})  of the  Large Binocular
Telescope (LBT).  According to the \citet[][]{Sand09} recovery of the
galaxy's star formation history (SFH),  Hercules is old ($>$ 12 Gyr,
with  negligible star formation in the last 12 Gyr) and metal-poor
([Fe/H]$ \sim - 2.0$ dex), with an intrinsic spread in metallicity and
both [Fe/H] = $-2.3$ dex and $-1.7$ dex populations contributing to
the SFH.  In the literature there are several spectroscopic and
photometric determinations of the metallicity  of Hercules. These
studies confirm that Hercules shows a spread in metal abundance,  with
values of the mean metallicity $\langle {\rm [Fe/H]} \rangle$ ranging
from about $-$2.0  to $-$2.7 dex
\citep{SG07,Kirby08,Koch08,Sand09,Aden09,Aden11}. 
 
The line of sight towards the Hercules dSph galaxy is heavily
contaminated by  Galactic foreground stars, making it hard  to
determine membership from the CMD alone. Even when radial velocities
are added the selection remains uncertain,  because the mean velocity
of the Hercules dSph galaxy coincides with the velocity of the thick
disk (\citealt{Aden09}).  Indeed,  in the \citet{Aden09}  CMD, based on
Str\"{o}mgren photometry   obtained with the Wide Field Camera of the
INT,  the RGB of Hercules is not easily identified due to the halo
foreground contamination. These authors used the  c1 index in the
Str\"{o}mgren system to disentangle the galaxy's RGB and HB stars  from
the foreground contamination.

Hercules appears to be highly elongated. The galaxy's structural
parameters were obtained by a number of different  authors
\citep{Bel07,Col07,Martin08,Sand09}  who, fitting different stellar
profiles to their independent photometric datasets, found similar
values for the galaxy's central position,  position angle ($\theta$),
and ellipticity ($\epsilon$), but rather different half-light radii.
In a recent analysis, \citet{Pen09} suggest that the density profiles
of relaxed, tidally stripped dwarf spheroidals like Hercules are
better  approximated by a Plummer law. Moreover, \citet{Sand09}
clearly point out the need for deep photometry in order to properly
constrain the structural parameters of the new faint MW satellites.
In the following, we shall adopt the structural
parameters obtained by \citet{Sand09} assuming a Plummer stellar
distribution, namely  $\theta=-72.59^{\circ}$, $\epsilon =0.67$ and
$r_h=6.27\arcmin$.  Hercules has  no evidence of internal
rotation, and  a very low velocity dispersion of $\sim 5$ km s$^{-1}$
for \citet{SG07} or 3.72 km s$^{-1}$ for \citet{Aden09}. The
explanation of such a large ellipticity in absence of a rotational
support might imply that Hercules is not in dynamical equilibrium  due
to  strong tidal distortions \citep[see discussion
in][]{Col07,Martin08}.   The latest studies estimate 
distances in the range of  $\sim 132$ kpc to $\sim 147$ kpc \citep[see
e.g.][]{Col07,Sand09,Aden11} and a total absolute magnitude ranging
from $M_V=-6.2\pm 0.4$ mag to $M_V=-6.6\pm0.3$ mag \citep[see
e.g.][]{Sand09,Martin08}. 

The study presented in this paper is based on $B,V$ photometric
time-series imaging covering a field of view (FOV) of $\sim 40'
\times 36'$ (see Table~\ref{t:obs}), extending  well
beyond Hercules' half-light radius. These data
have allowed us to obtain a complete inventory of the variable stars
belonging to  the galaxy and to trace the corresponding parent stellar
populations.

This paper is organized as follows:  in Section 2 we present the
observations and the data reduction procedures; Section 3 is devoted
to the variable stars, whereas the CMDs and the implications for the
structure of the Hercules UFD are discussed in Section 4. A new
estimate of the distance to Hercules based on the galaxy's RR Lyrae
stars is  presented in Section 5.  Finally, the summary and
conclusions in Section 6 close the paper.

\section{Observations and Data Reduction}

Time-series observations in $B$ and $V$ of the Hercules UFD were
obtained over the  period 2007 April to 2009 June, using a number of
different telescopes. The collected data and related
telescopes/instrumental set-ups are summarized in Table
\ref{t:obs},  where $N_B$ and $N_V$ are the number of frames in the
$B$ and $V$ bands, respectively. 

Pre-reduction of the images was performed by following standard
procedures (bias subtraction and flat-field correction) with
IRAF\footnote{IRAF is distributed by the  National Optical Astronomy
Observatories, which are operated by the Association of Universities
for Research in Astronomy, Inc., under cooperative agreement with the
National Science Foundation}, except for the WFI data, for which the 
pre-reduction was carried out using the VST-Tube pipeline \citep{Grado12}.
We then performed PSF photometry using
the DAOPHOT\,IV/ALLSTAR/ALLFRAME packages (\citealt{st87,st94}). After
an accurate evaluation of the PSF of each individual frame, a
reference image was built by averaging all the available frames and a
source catalogue was extracted from the stacked image. The source list
was then passed to ALLFRAME, in order to obtain a homogeneous
photometry of all images simultaneously, thus providing $b$ and $v$
instrumental magnitude catalogues for each telescope.  Typical
internal errors of the single-frame photometry for stars at the HB
magnitude level ($V \sim 21.3$ mag) are
of about 0.01 mag in both bands.

The absolute photometric calibration was derived using observations of
standard stars in the Landolt fields SA 101, SA 107, SA 110 and PG1323
\citep{la92,la09}, as extended by \citet{Stetson00}\footnote{see
http://cadcwwwdao.nrc.ca/standards.}, which were obtained at the INT
during the night of 2007 April 22. Errors of the absolute photometric
calibration are $\sigma_B=0.01$ and $\sigma_V=0.01$  mag.  As the
various telescopes we have used define slightly different photometric
systems, each individual dataset was tied to the INT standard
calibration independently.  The calibrated catalogues were then
combined using DAOMASTER \citep{st94} to create a final list
containing positions, robust intensity-weighted mean $B$ and $V$
magnitudes \citep{st94}, and weighted average values of the $\chi$ and
$Sharp$ parameters\footnote{In the single frame photometry file, the
$\chi$ parameter of each star is a robust estimate of the observed
pixel-to-pixel scatter of the fitting residuals to the expected
scatter, whereas the $Sharp$ parameter is related to the intrinsic
angular size of the astronomical objects, and for stellar objects
should have a value close to zero \citep{st87}.} for all objects. We
have adopted $ -0.4 < Sharp < 0.4$ and $\chi < 3$ to select the
stellar-like objects imaged in our total FOV of about 
$40' \times 36'$. This selection is reliable for magnitudes brighter than $V \sim$
24 mag, while the uncertainty increases at fainter magnitudes.

\section{Variable Stars}\label{s:variables}

Variable stars were identified from the $B$ and $V$ time-series
data, separately.  First, we calculated the Fourier transforms (in the
\citealt{sc96} formulation) of the stars having at least 12
measurements in each photometric band, then we averaged these
transforms to estimate the noise and calculated the signal-to-noise
ratios (S/Ns). Results from the $B$ and $V$ photometries were
cross-correlated, obtaining a list of 860 stars with $S/N >4$ in both
photometric bands (see e.g. \citealt{Musella09}).  A visual inspection
of the light curves allowed us to confirm the variability  of  10 of
these candidates. We also analyzed in detail  all stars falling on the
HB, in the region above the HB, where the Anomalous  Cepheids are
generally located, and in the Blue Straggler region, where one may
expect to find variables of the SX Phoenicis type. No further variable
stars were identified,  thus confirming the reliability of our
procedure to identify variable stars.  To study the light curves we
used the software Graphical Analyzer of Time Series (GRaTiS; see e.g.,
\citealt{clementini00}) that, beyond confirming the variability for
all the 10 candidates, also provided periods  accurate to 4-6 decimal
places for all of them.  The Hercules variables include 1 Anomalous
Cepheid (AC) and 9 RR Lyrae  stars, of which 6 are fundamental-mode
(RRab) and 3 are first-overtone (RRc) pulsators. 

\begin{figure}
\epsscale{.80}
\plotone{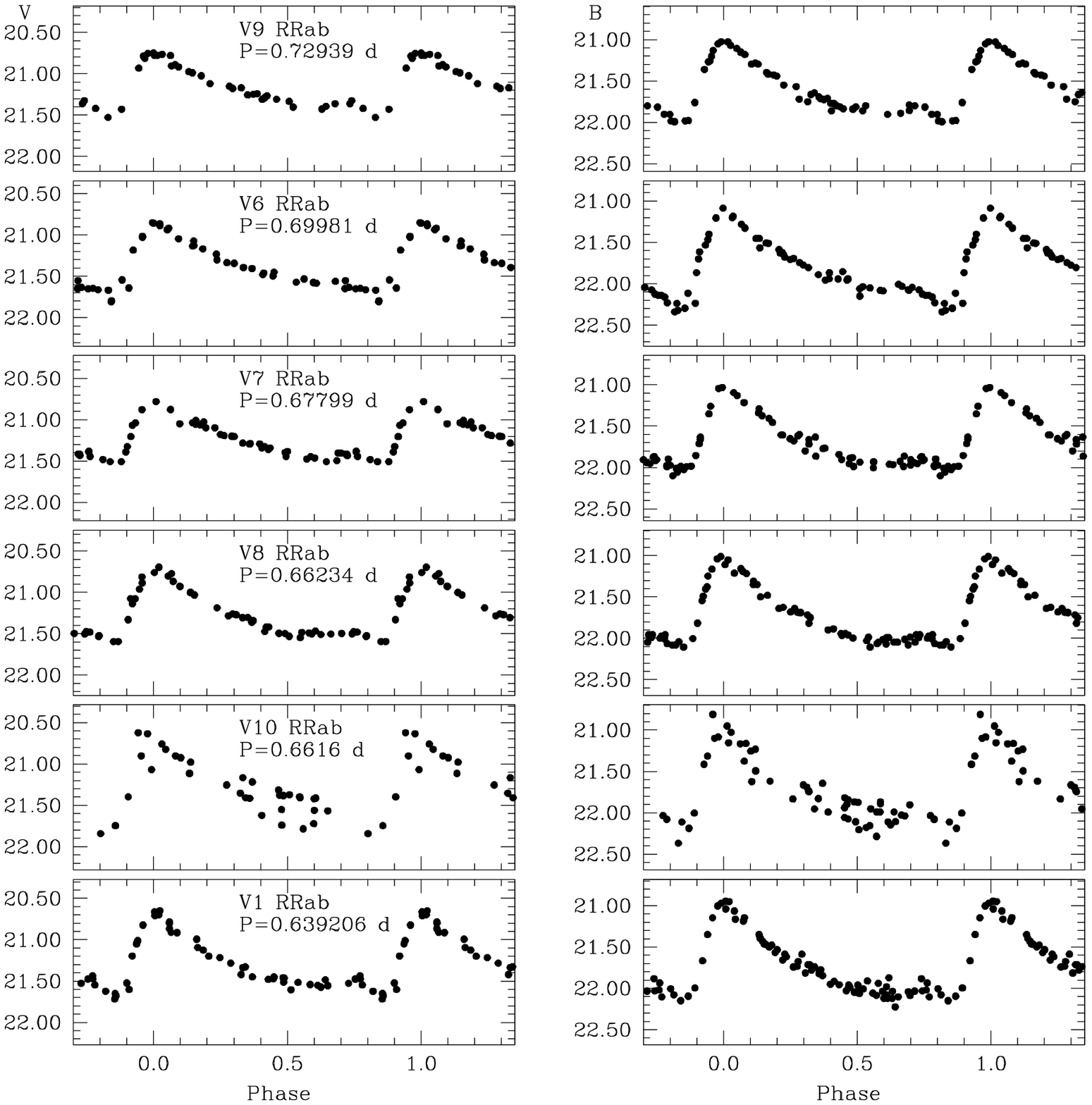}
\caption{Light curves of Hercules fundamental mode RR Lyrae stars. Variables are ordered by decreasing period. Typical
  instrumental errors of the single data points  
are of  $\sim 0.01$ mag for $V$ in the range of 20.0 to 22.0 mag, and $\leq 0.01$ mag for $B$ in the range of  20.0 to 22.0 mag. 
}\label{f:fig1}
\end{figure}

\begin{figure}
\epsscale{.80}
\plotone{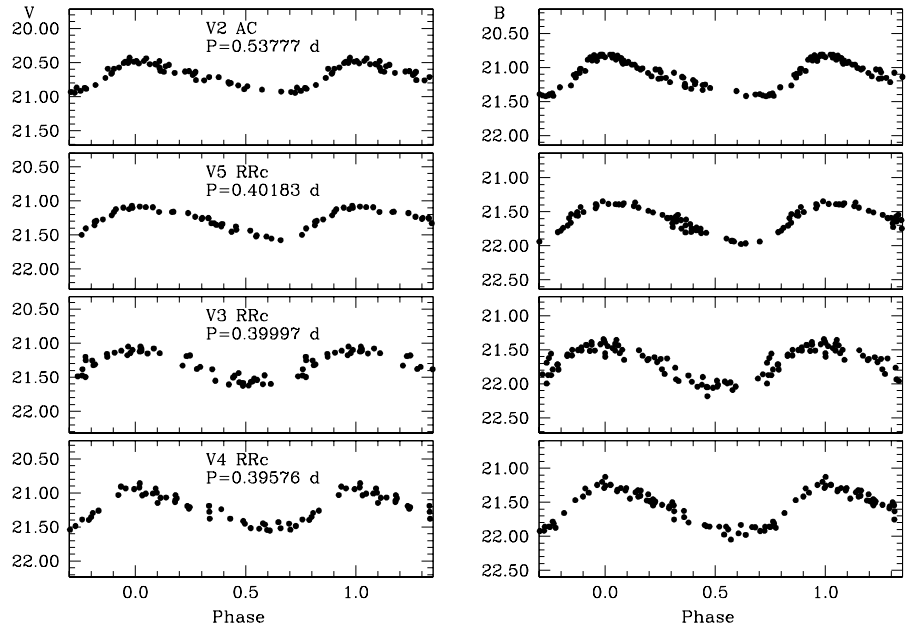}
\caption{Light curves of the Hercules AC (top panel) and first overtone RR
  Lyrae stars. Variables are ordered by decreasing period. Typical
  instrumental errors of the single data points  
are of  $\sim 0.01$ mag for $V$ in the range of 20.0 to 22.0 mag, and $\leq 0.01$ mag for $B$ in the range of  20.0 to 22.0 mag. 
}\label{f:fig2}
\end{figure}

Classification and properties of the confirmed variable stars are
summarized in Table~\ref{t:RR}, whereas their light curves are shown in
Figs.~\ref{f:fig1} and \ref{f:fig2}. We have assigned to the variables
increasing numbers starting from the galaxy center that was set at
R.A.$=16^h 31^m 03.12^s$, Decl.$=12^{\circ}47^{\prime}14.01^{"}$,
J2000.0  \citep[][]{Sand09}.  Time-series data for the variable
stars are provided in Table~\ref{t:pho}.  The light curves  are very
well sampled, and show very little scatter, except for star V10.   The
star does not appear to be blended with other stars on the images.  We
suspect that V10 might be affected by the Blazhko effect
\citep{Blazhko1907}.

We note that, according to the period, the AC (star V2)  could  as
well be  a fundamental-mode RR Lyrae star. However, the star's average
magnitude is about 0.4-0.5 mag brighter than the HB level.  This
occurrence,  together with a too small amplitude for the star's short
period, suggest that the variable either suffers from blending by a
contaminant  star or is, perhaps, an AC. The photometric parameters
$\chi$ (goodness of the PSF fitting) and $Sharp$ (estimate of the
object shape) provided for V2 by the ALLFRAME package seem to rule out
the presence of detectable companions around the star, thus favoring
the  AC hypothesis.  Another possibility is that V2 could be an
overluminous RRc as, for instance, V70 in the M3 GGC. However,  the
period of V2 (0.53777 d)  is  definitely  longer than that of M3-V70
($P=0.4865$ d, \citealt{Kaluzny98}, \citealt{Carretta98}).  On the
other hand, the mean magnitude and the period of V2 are in very good
agreement with the AC period-luminosity relations by \citet[][and
references therein]{BW2002}, \citet{Marconi04} and \citet[][and
references therein]{Baldacci04}, thus favoring an AC classification
for V2.  Finally,  as described in Section \ref{s:cmd}, the AC
classification is further  supported by  the comparison with
evolutionary tracks.

The AC,  all the RRc variables and three of the RRab stars lie inside
the half-light radius of the Hercules galaxy. One RRab is just on the border of this region
and the remaining two RRab stars lie well
outside the half-light radius, with the farthest from the center being confirmed
as a member by \citet{Aden09}  on the basis of both  radial velocity and
Str\"{o}mgren photometry (see Section \ref{s:cmd} and
Figs. \ref{f:cmd} and \ref{f:ad}
for details).  All the RR Lyrae stars fall on the HB of the Hercules
UFD,  and their mean magnitude is consistent, within the errors, with
the average luminosity of  the HB  inferred  by fitting the galaxy CMD
with  the ridge line of the GGC M68 (see section \ref{s:cmd} for
details). 

\begin{figure}
\epsscale{.80}
\plotone{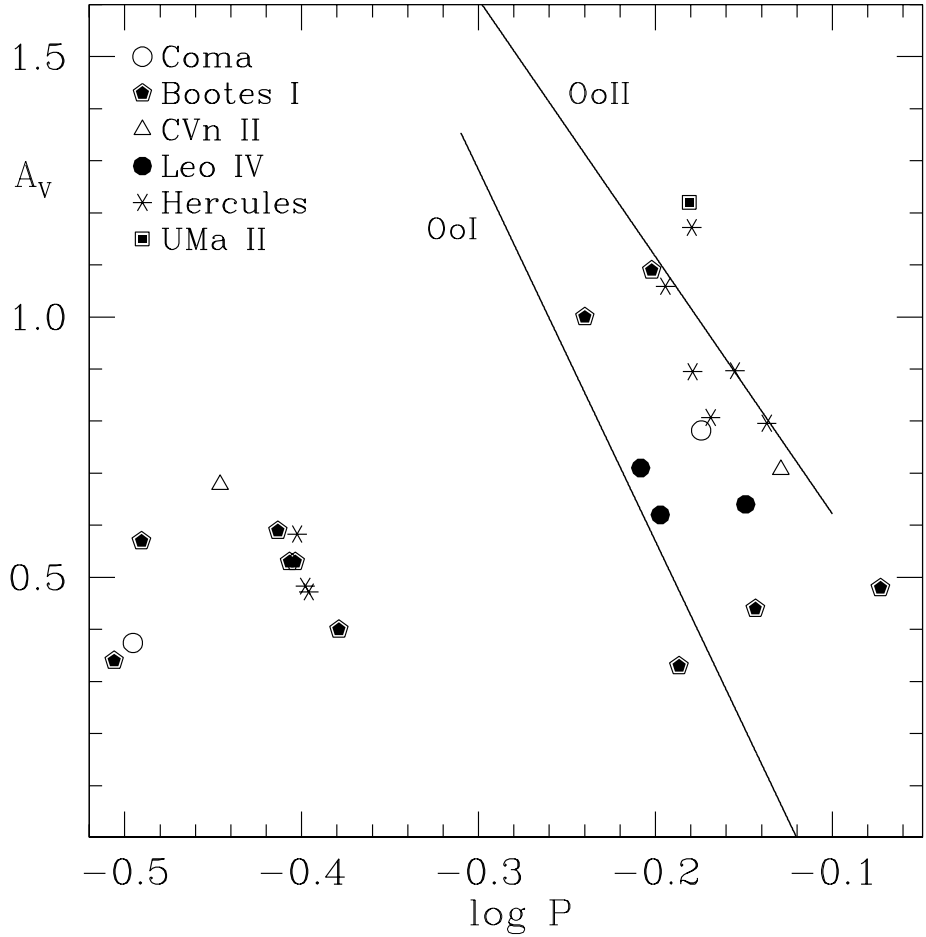}
\caption{$V$-band period-amplitude diagram of the RR Lyrae stars in the UFDs
  studied so far for variability:  Coma,
Boo I, CVn II, Leo IV, UMa II and Hercules UFDs. Solid
lines are the positions of the Oo I and II Galactic GCs, from \citet{CR00}.}
\label{f:bailey}
\end{figure}

The mean period of the  RRab stars is $\langle P \rangle=0.68$ d ($\sigma$=0.03 d), 
thus classifying Hercules as an Oosterhoff type-II system.
Fig.~\ref{f:bailey} shows the position of the Hercules RR Lyrae stars
on the $V$-band period-amplitude (Bailey) diagram. We have also
reported the RR Lyrae stars identified  in the other MW UFDs we have
studied so far, for comparison. Hercules' RRab stars (star symbols)
lie closer to the loci of the Oo~II systems \citep[from][solid
line]{CR00} and the RRc stars fall on the long-period tail of the
bell-shaped distribution defined by  RRc stars in Oo~II systems 
(see, e.g., Fig. 9 of \citealt{dicriscienzo2011}, and Fig. 4 of \citealt{cacciari05}).

\subsection{Metallicity and reddening from the RR Lyrae stars}\label{s:met}

As shown by \citet{JK96}, it is possible to use the  shape of an RR
Lyrae $V$-band light curve to obtain an estimate of the star's
metallicity. We have performed a Fourier decomposition of the
$V$-band light curves of the variables in Hercules and calculated
amplitude ratios $A_{n1}$ and  phase differences $\phi_{n1}$. We list
in Table \ref{t:fou} these quantities up to the term $n=3$. The
Fourier parameters were then used along with the formulae provided by
\citet{JK96} and \citet{Morgan07} for ab- and c-type RR Lyrae stars,
respectively,  to obtain individual metallicities for the Hercules  RR
Lyrae stars, except for V10 whose light curve is too noisy for a
reliable application of the Jurcsik \& Kovacs method.  The metallicity
estimates obtained with this technique are reported in column 2 of
Table~\ref{t:met}; they  are on the \citet{ZW84} scale.   These
individual metallicities were transformed to the \citet{Carretta09}
metallicity  scale (using the \citealt{Kapakos11} transformations for the ab-type variables,
 and the  \citealt{Carretta09} recalibration for the c-type RR Lyrae stars, see Column 3 of
Table \ref{t:met}). We then
averaged these values (weighted average) to obtain our best estimate for the
metallicity of Hercules' old population. Our resultant metallicity is 
$\langle {\rm [Fe/H]_{RR,C09}} \rangle= -2.30 \pm
0.15$ dex. 

 Literature values for the Hercules UFD's metal abundance are  summarized in
Table \ref{t:lit_met}. They were obtained using a variety of
different methods and span a rather large range. In particular,  we
note that the values from \citet{Col07} and \citet{Sand09}  are photometric estimates
based on fitting the galaxy's CMD with evolutionary tracks,  while all
other entries in the table are spectroscopic determinations.
Specifically, the \citet{Koch08} value is based on abundance analysis of
medium-high resolution ($R\sim 20,000$) spectra of two stars in
Hercules.  The \citet{SG07} analysis is based on medium-resolution spectra
of 30 red giants in Hercules, and the measurement of  the equivalent
widths  of the  Ca triplet absorption lines.  The \citet{Kirby08}
study is a
re-analysis of the \citet{SG07} spectroscopic data, based on an automated
spectral synthesis technique.  This latter method likely provides
systematically lower metallicities (by $\sim 0.15$ dex), as discussed
by the authors themselves \citep[see][]{Kirby09,Kirby10}. 

Finally, in a recent study,  \citet{Aden11} confirm that a large
metallicity spread ($-3.2 \leq {\rm [Fe/H]}\leq -2.0$ dex,  for an
average value of [Fe/H]=$-2.70\pm 0.38$ dex) exists in
Hercules, from the  analysis of medium-high resolution ($R\sim
20,000$) spectra  of  11 RGB stars.  However,
taking into account only the  red giants with metallicity measurements
based on a significant number of iron lines ($\ge 5$; four stars
according to Table~6 in \citealt{Aden11}), we obtain  $\langle
{\rm [Fe/H]}\rangle=-2.3 \pm 0.2$ dex (and a range of  
($-2.4 \leq {\rm [Fe/H]}\leq -2.0$ dex).    This value is in excellent
agreement with our estimate from the RR Lyrae stars.
On this basis, we will adopt  the
metallicity estimate $\langle [Fe/H]_{RR,C09} \rangle$ obtained from the RR Lyrae
stars  in the following analysis.

Hercules' reddening can be estimated from  the galaxy's  RR Lyrae stars.
Using the relation for RRab stars by \citet{P02},  that 
connects the color excess to the star's B-band amplitude,
the logarithm of the period and the metallicity, we obtain a mean
color excess  $E(B-V)_{RRab}=0.09 \pm 0.02$ mag, where  we have
adopted the individual metallicities on the \citet{Carretta09} scale
listed in Table~\ref{t:met}.  We also use the method by \citet{Sturch1966}, 
which is based on the RR Lyrae's $B-V$ color at minimum light,  and
obtain  $E(B-V)_{RRab} = 0.12 \pm 0.03$ mag, by applying  the
metallicity-dependent relation by \citet{Clementini03}. This estimate
is larger than the previous value but, as already pointed out by
\citet{Walker98} and \citet{Clementini03},  Sturch's method
overestimates the color excess by 0.01-0.03 mag.  We therefore 
adopt $E(B-V)=0.09 \pm 0.02$ mag for the reddening.
This is in very good agreement with the value of $0.084 \pm 0.026$ mag
derived from the \citet{Schlegel98} maps.

\section{CMD and structure of Hercules} \label{s:cmd}

\begin{figure}
\epsscale{.80}
\plotone{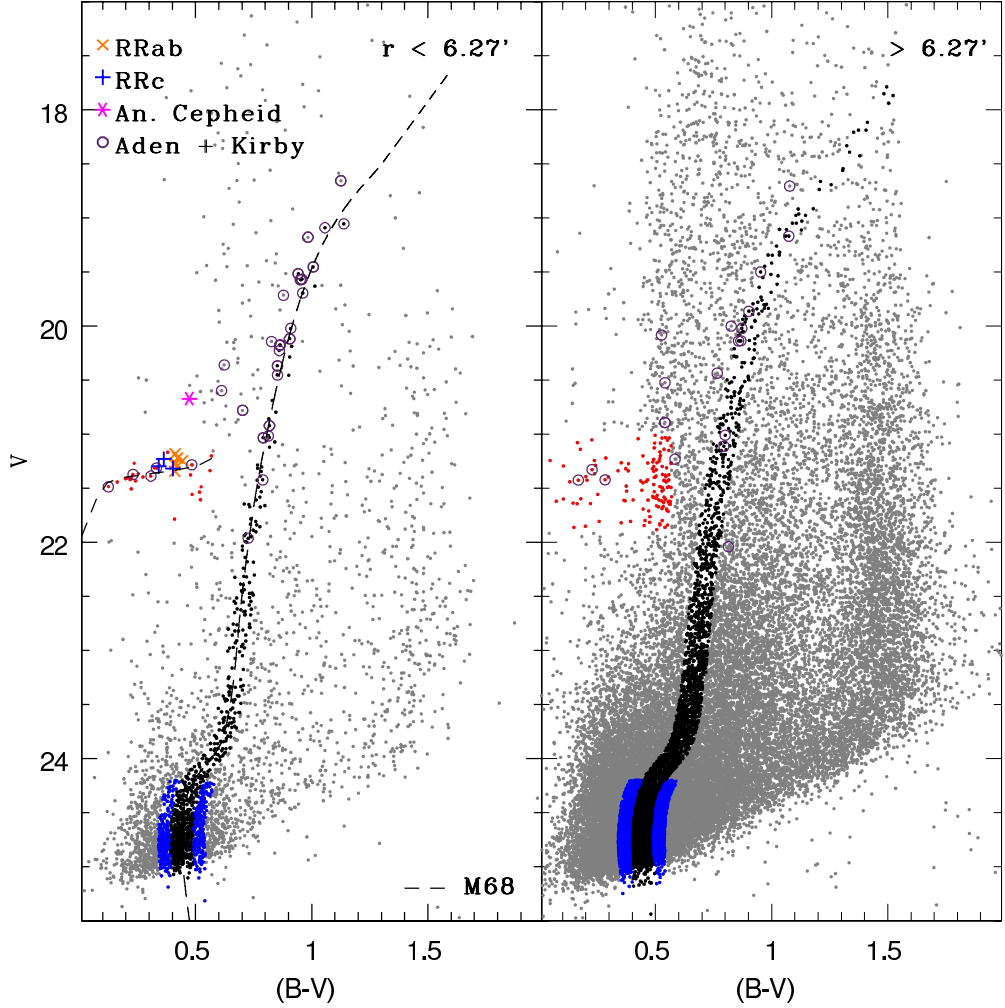}
\caption{Left: 
 $V, B-V$ CMD of the Hercules UFD, drawn from all the stellar-like
objects  (see text for details) within the galaxy half-light radius
($r_h \simeq 6.27\arcmin$, \citealt{Sand09}).  The dashed (black) line is
the ridgeline of the GGC M 68.  Black and blue dots are stars
respectively within $\pm 0.05$ mag in $B - V$ and from $\pm0.05$ to
$\pm 0.1$ mag from the ridgelines of M 68.  Red dots are non-variable
stars in the HB region.  The RRab and RRc stars are marked by orange crosses
and blue plus signs,  respectively,  the Anomalous Cepheid by a
magenta asterisk. Open violet circles mark member stars of the Hercules UFD
identified spectroscopically by \citet{Kirby08} and
spectro-photometrically by \citet{Aden09} (see text for details). {\it
Right:} $V, B-V$ CMD of all the stellar-like objects outside the half
light radius.  (Symbols and color-coding are as in the left panel).  }\label{f:cmd}

\end{figure}

Fig. \ref{f:cmd} shows the $V$, $B-V$ CMD of Hercules obtained in the
present study.  In the left panel we plot stellar-like objects within
the galaxy's half-light radius; in the right panel, instead, we show
objects outside this region over our FOV of $40' \times 36'$.  
All the RR Lyrae stars have been plotted
in the left panel of the figure including those falling outside the
galaxy half-light radius. The CMD reaches $V \sim $ 25 mag and
appears to be heavily contaminated at each magnitude level by field
objects belonging to the MW halo and disk, as well as by background
galaxies. Our photometry reaches a few magnitudes deeper than the SDSS
photometry (see Fig. 2 of  \citealt{Bel07}); while it is roughly
1 mag shallower than the photometry obtained
using the LBT (Fig. 1 in \citealt{Col07},  and Fig. 8 in
\citealt{Sand09}). However, within the errors, it is consistent both in $B$ and $V$, with the
photometry reported in Table 2 of \citet{Sand09}.

In Fig. \ref{f:cmd}, the main branches of the Hercules CMD are barely
distinguishable due to the overwhelming  contamination by the MW
field.  To identify  stars belonging to  Hercules  we used the method
that we have already successfully applied  in our previous papers (see
e.g., \citealt{Musella09,Moretti09}). 
Specifically, we used the mean ridgelines of the GGC M68 (dashed
black lines in Figure \ref{f:cmd}), obtained from \citet{Walker94} $B, 
V$  photometry to fit the HB and RGB of the Hercules UFD, by letting
the M68 ridgelines vary within the values of the cluster reddening and
distance modulus available in the literature.  The ``by eye'' best fit
was obtained for a shift of $\Delta V = +5.68$ mag in magnitude and
$\Delta (B - V ) = +0.02$ mag in color. We adopted   
M68 for identifying members of the Hercules UFD
because, like Hercules, it is very metal-poor.  Its metallicity,
${\rm [Fe/H]_{M68}} = -2.27 \pm 0.04$ dex \citep[][]{Carretta09}, is in very
good agreement with the mean metallicity obtained in section
\ref{s:met} from the Hercules RR Lyrae stars. Furthermore, the cluster has a
well-defined and tight RGB, as well as an extended HB including stars
both redder and bluer than the RR Lyrae instability strip
(\citealt{Walker94}), thus  resulting better suited to identify the galaxy's HB than, 
for instance, M92, which instead was used as a fiducial by  
\citet{Bel07}.
With this procedure we selected as most
probable members of the Hercules galaxy the sources lying within
$\pm0.05$ mag in $B-V$ from the ridgelines of M68 (black dots in the
left panel of Figure \ref{f:cmd}). To account for the larger
photometric errors, below $V = 24.2$ mag we extended this range within
$\pm 0.10$ mag of the ridgeline of M68 (blue dots)
Adopting for M68
a reddening value of $E(B - V ) = 0.07\pm 0.01$ mag
\citep[][]{Walker94} the color shifts needed to match the HB and RGB
of Hercules imply a reddening of $E(B - V ) = 0.09 \pm 0.01$ mag for
the galaxy, in excellent agreement with the value  obtained from the
RRab stars (see Section \ref{s:met}). 
Totally similar results are obtained  using the ridgelines of
the GGC M15 that is slightly more metal-poor than M68 and matches
equally well the main branches of the Hercules CMD, whereas the
ridgeline of the metal-intermediate GC M3
\citep[NGC~5272;][]{Ferraro97,JB98} is too red and would require a
negative reddening to match the galaxy RGB.

With the help of M68 ridgelines, it is possible to determine the
average luminosity of the Hercules  HB in the region of the so-called RR
Lyrae gap ($\langle V_{HB} \rangle = 21.35 \pm 0.03$ mag), and to locate the galaxy's 
 main-sequence turnoff at $V \sim 24.4$ mag.

Our identification of  Hercules members 
is supported by the spectroscopic study of \citet[][20 stars]{Kirby08} and 
the spectrophotometric analysis of  \citet[][47 stars]{Aden09},  with membership to
the Hercules UFD confirmed by radial velocity measurements and,  for
\citet{Aden09}'s  sample, also by Str\"{o}mgren photometry (violet
open circles in Fig. \ref{f:cmd}). The excellent 
agreement between these studies and our results supports
the reliability of the procedure we have used to select Hercules
members and to identify the HB. 

\begin{figure}
\epsscale{.80}
\plotone{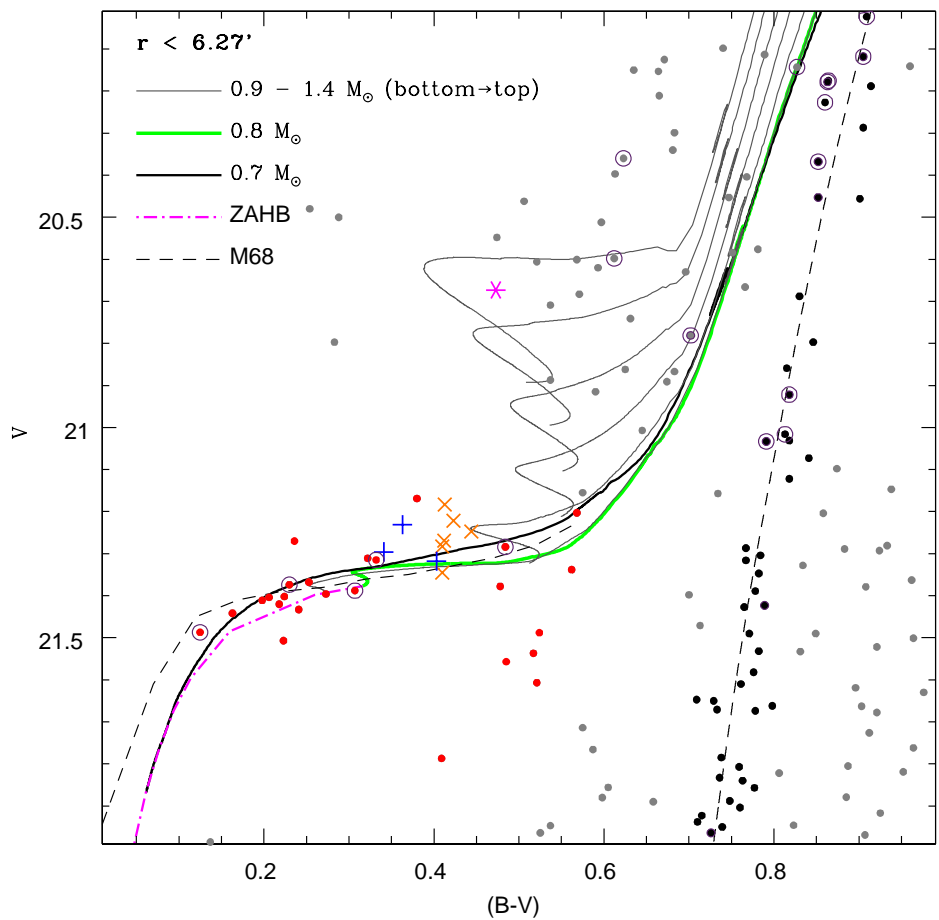}
\caption
{The thick black and green lines are the HB evolutionary tracks
for 0.7 and 0.8 $M_{\odot}$, respectively, whereas the thin grey
lines are for masses from 0.7 to 1.4 $M_{\odot}$ (from bottom to top, 
with step of 0.1 $M_{\odot}$) including the range of masses
corresponding to the HB turnover \citep{CD95}.  The magenta
dot-dashed line represents the Zero Age Horizontal Branch (ZAHB)  for the
same chemical composition.  The black dashed lines are the ridgelines of the GGC M68. }\label{f:cmd_tracce}

\end{figure}

\citet{Bel07} compared the galaxy's CMD with the ridge-line of
the GGC M92  ($[Fe/H] \sim -2.24$ dex) and M13 ($[Fe/H] \sim -1.65$
dex) and concluded that the morphological features of  Hercules CMD
are generally well described by the ridge line of the old, metal-poor
globular cluster M92.  Using suitable metal-poor evolutionary tracks
and their deep LBT photometry to recover the galaxy's SFH,
\citet{Col07} and \citet{Sand09} find that Hercules hosts a single old
stellar population ($> 12$ Gyr),  and shows no evidence of significant
subsequent star formation events.  This is in contrast with  the
\citet{Bel07} claim that Hercules' extended HB showing both  a blue
and a red component  might be the signature of possible multiple
stellar populations in the galaxy.  However,   \citet{Bel07}'s
argument cannot  be considered conclusive, as  the morphology of the
HB is  driven by   a complex interplay of different effects,
traditionally known in the literature related to GGCs as the ``second
parameter problem'' \citep[see e.g.][and references
therein]{Lee94,Buonanno97,Cat09,Gratton10}. On the other hand, in the
region of the CMD above the HB ($V \sim 20.5$  and $(B-V) \sim 0.5$
mag), there seems to be an overabundance of stars that might represent
the signature of an intermediate-age population in
Hercules. Unfortunately,  the contamination  by field stars in this
region of the CMD is very high (see right panel of Fig. \ref{f:cmd}),
thus we  cannot reach any firm conclusion with our data.  An analysis
of the existent literature data  shows that  the HB and the brighter
portions of Hercules' CMD are very poorly populated in the  LBT
photometry of \citet{Col07}  and \citet{Sand09}, likely due to
saturation effects, and cannot be used to check the possible existence
of such an excess above the HB. However,  an overabundance of stars
brighter than the HB was observed in Hercules also by \citet{Aden09},
who investigated whether  they might be variables, based on the
magnitude variation  in their few (no more than 3) epoch data for the
stars.  They provide a list of possible variable stars in their Table
7.  Though we note that these candidate variables are not included in
the \citet{Aden09} final list of Hercules members, we checked them in
our time series photometry,  and do not confirm their variability.
However,  we have identified an  AC variable in this region above the
HB,  star V2  (asterisk in Fig. \ref{f:cmd}, see Section
\ref{s:variables}).  In Fig. \ref{f:cmd_tracce} we show an enlargement
of the HB region of the CMD in Fig. \ref{f:cmd}  (with the same color
coding), where we have plotted  the helium burning evolutionary tracks
of the stellar model database BaSTI\footnote{The BaSTI database is
available at: http://www.oa-teramo.inaf.it/BASTI}
\citep{Pietri04,Pietri06}, for stellar masses in the range of 0.7 to
1.4 $M_{\odot}$ (with a step of 0.1 $M_{\odot}$) and metal abundance
$Z=0.0001$.  In particular,  the thick (black and green) lines
are the HB evolutionary tracks for 0.7 and 0.8 $M_{\odot}$,
respectively, whereas the thin (grey) lines are for $M \ge 0.9
M_{\odot}$ including the range of masses corresponding to the HB
turnover \citep{CD95}.  The magenta dot-dashed line represents
the Zero Age Horizontal Branch (ZAHB)  for the same chemical
composition. The location of the RR Lyrae stars is consistent with
the model predictions and suggests an evolutionary effect for some of
them. On the other hand, the AC, star V2,  is, as expected, consistent
with the evolution of a $M \sim 1.35 M_{\odot}$ from the turnover
region of the ZAHB.  The origin of ACs is still debated \citep[see
e.g.][and references therein]{Marconi04} and the most widely accepted
interpretations are: (1) they are young ($\leq 5$ Gyr) single stars
due to recent star formation; (2) they formed from mass transfer in
binary systems as old as the other stars in the same stellar
system. Our analysis does not allow us to  discriminate between these
two scenarios, however, we note that, in the first hypothesis,  the
comparison of V2 with the evolutionary tracks would suggest   the
existence in Hercules  of an intermediate-age population of stars as
old as $\sim 2-3$ Gyrs.

\begin{figure*}
\epsscale{.80}
\plotone{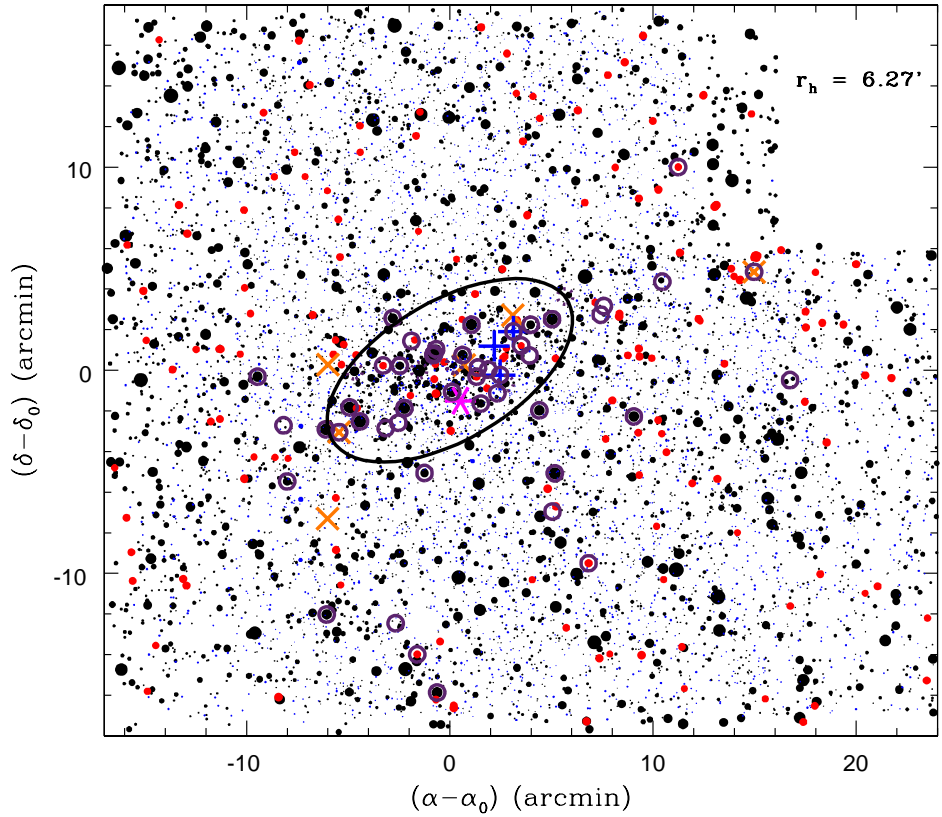}
\caption{Map of the stars we consider to belong
to the Hercules galaxy in the FOV ($\sim 40' \times 36')$ of our observations.
Symbols and color coding are the same as in Fig. \ref{f:cmd}. Symbol
sizes are inversely proportional to the object's magnitudes. A  black ellipse
describes the half-light region of the galaxy, for the angle position  and 
ellipticity obtained for Hercules by \citet{Sand09}.}\label{f:ad}
\end{figure*}

\begin{figure*}
\epsscale{.80}
\plotone{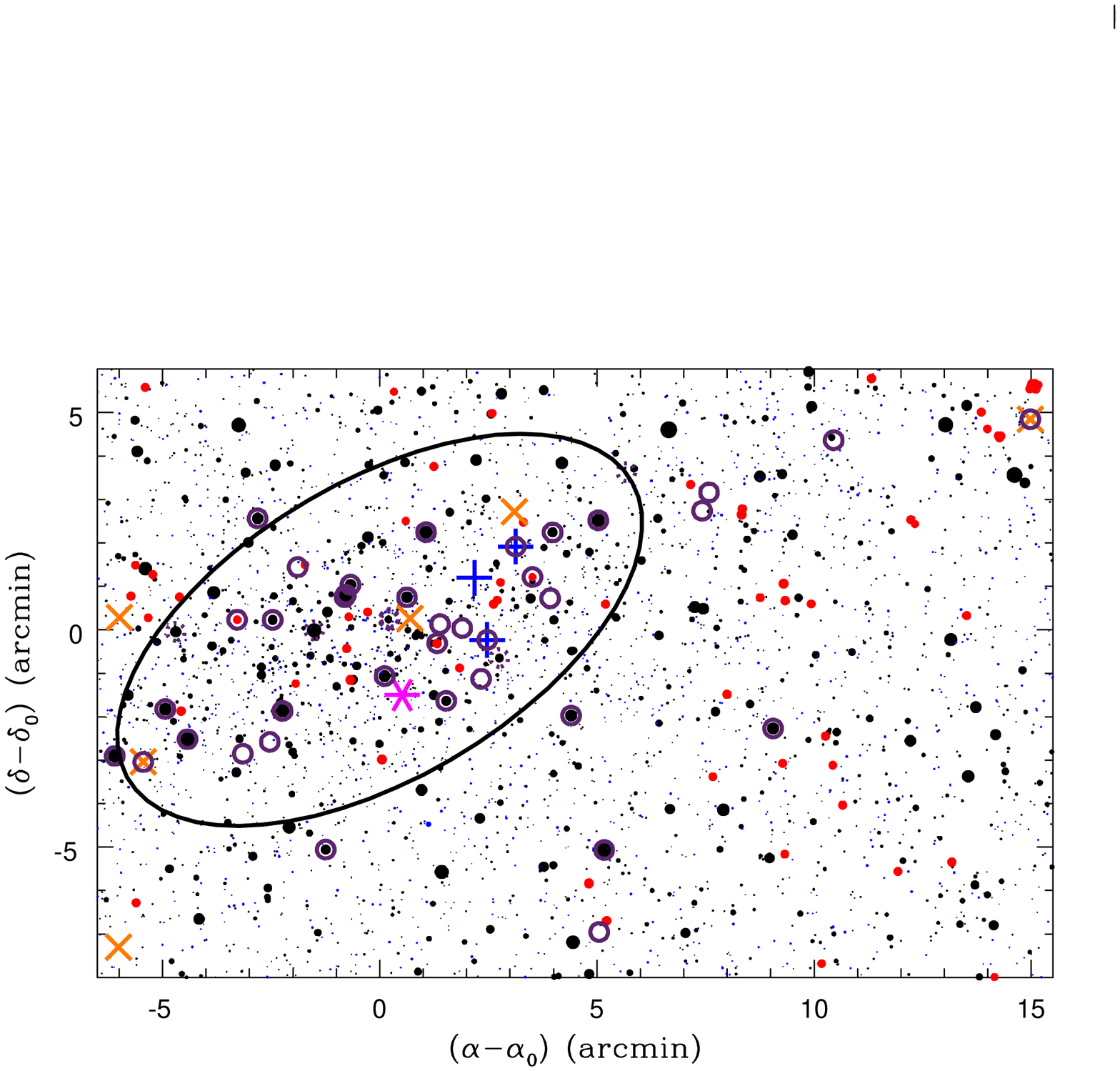}
\caption{Zoom of the map in Fig. \ref{f:ad} with the galaxy  
 half-light radius region and all the variable stars identified
  in this work.}\label{f:adzoom}
\end{figure*}

Fig. \ref{f:ad}  shows the position of the stars we consider to be
members of the Hercules galaxy in our FOV.  Symbols and color-coding
are the same as in Fig. \ref{f:cmd},  and the symbol size is inversely proportional
to the object's magnitude. The black ellipse corresponds to the
half-light  radius, the  angle  position,  and the ellipticity obtained for
Hercules by \citet{Sand09}.  This map confirms that  the galaxy is
elongated and has an irregular and extended shape. Likely, Hercules
was   disrupted due to tidal interaction  and is now embedded in the
MW halo.  Our discovery of two RRab variables outside the half-light
radius, together with the spectrophotometric identification in the
same external region of some stars with confirmed membership by
\citet{Aden09}, support the tidal interaction scenario. 
  Fig. \ref{f:adzoom} shows an enlargement of the map in Fig. \ref{f:ad} with  the galaxy half-light 
radius region and all the variable stars identified in the present work.

\section{A new estimate of the distance to Hercules}\label{s:distance}

The RR Lyrae stars  we have detected in Hercules  give us  the opportunity to
 estimate the distance to the galaxy using these  variables as standard candles. 

The position of the Hercules RR Lyrae stars in the CMD is in
satisfactory agreement with the  fiducial-line HB of M68, although
with a significant spread. The most deviating (brightest) of the RR
Lyrae variables, star V9, is located well outside the galaxy's
half-light radius (see Figs. \ref{f:ad}  and \ref{f:adzoom}).  Thus,
its high luminosity could be caused by a projection effect.  The
average magnitude of the  RR Lyrae stars  lying inside the galaxy's
half-light  radius is $\langle V_{RR} \rangle = 21.28 \pm 0.05$ mag
(average on 7 stars), where the error is the standard deviation of the
mean.  To obtain the absolute visual magnitude of Hercules' HB stars,
we adopt for the slope of the RR Lyrae magnitude-metallicity
calibration the value ${\rm \Delta M_V /\Delta [Fe/H]} = 0.214 (\pm
0.047)$ mag dex$^{-1}$ \citep[][]{Clementini03}, and consider two
different calibrations of the zero point.  At [Fe/H]=$-1.5$ dex,  we
adopt: ${\rm M_V=0.54} \pm 0.09$ mag, that is based on the Large
Magellanic Cloud  distance modulus of 18.52 $\pm 0.09$ mag of
\citet[][]{Clementini03}, and, alternatively,    ${\rm M_V}=0.45 \pm
0.05$ mag,  from \citet{Ben11}.   Adopting for  metallicity and
reddening the values derived in Section \ref{s:met}  ($\langle {\rm
[Fe/H]_{RR,C09}\rangle}= -2.30 \pm 0.15$ dex, and $E(B-V)_{RRab}=0.09
\pm 0.02$ mag, respectively)  we obtain true distance moduli of
$\mu_{0,[C03]}=20.6\pm 0.1$ mag ($D_{[C03]}=132 \pm 6$ kpc),  and
$\mu_{0,[Ben11]}=20.7 \pm 0.1$  mag ($D_{[Ben11]}=138 \pm 6$ kpc), for
the \citet{Clementini03} and \citet{Ben11} calibrations,
respectively.  The errors on the distance values include the
contribution of the uncertainties on the metallicity, the reddening,
the adopted slope of the $M_V -{\rm [Fe/H]}$ relation,  and the
average apparent visual magnitude of the Hercules RR Lyrae stars.

An additional estimate of the distance,  based on the RR Lyrae stars,
can be obtained using the theoretical Wesenheit relation in the $B,V$
bands, as defined by \citet{Dicrisci04}, with the assumption of a
suitable evolutionary mass for  the metallicity of Hercules RR Lyrae
stars. The resulting distance modulus of $20.6 \pm 0.1$ mag is in
perfect agreement with $\mu_{0,[C03]}$. Moreover,  a distance of $132
\pm 6$ kpc is in excellent agreement with the estimates by
\citet{Col07}, \citet{Martin08}, \citet{Sand09}, and in statistical
agreement with the distance by \citet{Bel07}. On the other hand, the
distance $D_{[Ben11]}$ appears to be only in statistical agreement
with all previous determinations.  To conclude this discussion, we
note that the very long estimate of $147^{+8}_{-7}$ kpc, obtained by
\citet{Aden09}  is likely  due to their HB mean magnitude being based
on a few stars, of which some are variables observed at random phase,
as we have confirmed in our study. 

Finally, the apparent distance modulus adopted in Fig. \ref{f:cmd_tracce},
$\mu=21$ mag,  corresponds to a true distance modulus of 20.7 mag.
However,  we should take into account the theoretical uncertainties on
the luminosity of the ZAHB and, in particular,  the remark by
\citet{Cassisi07} that, adopting updated conductive opacities, might
cause an increase in the ${\rm M_V(ZAHB)}$ of  about 0.06 mag at the
metallicity of Hercules, and a corresponding decrease in the inferred
distance modulus.  On this basis,  and considering also the results of
the theoretical Wesenheit relation, we thus conclude that our most
reliable estimate for the  distance modulus of Hercules is:
$\mu_0=20.6\pm0.1$ mag.

\section{SUMMARY AND CONCLUSIONS}

In this paper we have presented the first time-series analysis  of the
Hercules UFD. Using a variety of telescope/instrument facilities we
secured $\sim$ 80 and 50 epochs in $B$ and $V$.  These
data allowed us to detect and characterize  9 RR Lyrae stars (6 ab-
and 3 c-type, respectively) and one Anomalous Cepheid. The same
observations allowed us to build a deep CMD  extending well beyond the
galaxy's half-light radius.  The main results of this study are listed
below:
\begin{itemize}
\item The average period of the ab-type RR Lyrae stars,
$\langle$P$_{ab}\rangle=$0.68 d, qualifies Hercules as an Oosterhoff II  system,
in good agreement with the vast majority of the UFDs investigated so
far.  This occurrence favors the hypothesis that the UFDs could be the
``building  blocks'' of the Galactic halo, since the pulsation
characteristics of their RR Lyrae stars  are in agreement with the
properties of the  MW halo  variables.  
\item Hercules' CMD is dominated by a stellar population at least as
old and metal-poor as the GGC M68. This result is in
agreement with previous findings.  The HB shows some spread. We also
detected an overabundance  of stars above the HB, thus  confirming the
previous finding by \citet{Aden09}. This, along with the detection  of
an  Anomalous Cepheid  very likely belonging to Hercules,  hints at
the possible 
presence of an intermediate-age population about  $\sim 2-3$ Gyrs old
in Hercules.
\item The spatial distribution of Hercules' stars confirms the
elongated shape of this galaxy. The signature that Hercules is
undergoing tidal  disruption is provided by  the absence of a clearcut
difference between galaxy and field star properties,   and by the
presence of  two RR Lyrae stars lying well beyond the galaxy's
half-light radius. 
\item The RR Lyrae variables  were used to obtain  independent
estimates of  the metallicity, reddening and distance to Hercules, for
which we find [Fe/H]=$-2.30\pm0.15$ dex, $E(B-V)=0.09\pm0.02$ mag,
and $(m-M)_0=20.6\pm0.1$ mag respectively,  in very good agreement with literature
values. 
\end{itemize}

\bigskip 

\acknowledgments   
We thank an anonymous referee for carefully reading the paper and for providing comments that helped to improve the clarity of the 
manuscript. Financial support
for this research was provided by COFIS ASI-INAF I/016/07/0, by  the
agreement ASI-INAF I/009/10/0, and by PRIN INAF 2010, (P.I.:
G. Clementini). Support for M.C. is provided by Proyecto Fondecyt
Regular \#1110326; BASAL Center for Astrophysics and Associated
Technologies (PFB-06); FONDAP Center for Astrophysics (15010003); the
Chilean Ministry for the Economy, Development, and Tourism's Programa
Iniciativa Cient\'{i}fica Milenio through grant P07-021-F, awarded to
The Milky Way Millennium Nucleus; and Proyecto Anillo ACT-86. HAS
thanks the U.S. NSF for support under grants AST0607249 and
AST0707756.

{\it Facilities:} \facility{ING:Herschel}, \facility{ING:Newton}, \facility{Liverpool:2m}, \facility{FTN}, \facility{Max Planck:2.2m}



\clearpage



\clearpage

\begin{table}
     \caption[]{Instrumental set-up and log of the observations of Hercules} 
	 \label{t:obs}
	 \scriptsize
	\begin{tabular}{llllcrr}
	   \hline
	    \hline
	   \noalign{\smallskip}
  {\rm~~~~Dates}& { \rm Telescope/Instrument} & {\rm    ~~~Detector}&{\rm ~Resolution}  & { \rm FOV} &{\rm N$_B$}&{\rm  N$_V$}\\
 	        	           &                                   &~~~~           {\rm (pixel)} & {\rm  ~~($\prime\prime$                                  /pixel)}    	        &    & 	 \\
	    \noalign{\smallskip}
	    \hline
	    \noalign{\smallskip}
{ Apr. 2007-May2008}  & { 2.5m INT/Wide Field Camera}	   & { 4 of 4k $\times$ 2k}&~~~0.33 &~34' $\times$ 34' & 52$\ $& 32$\ $\\
{ May 2007}  & { 4.2m WHT/Prime Focus Camera}	   & { 2 of
  2k $\times$ 4k}	       &~~~0.24 &~16.2' $\times$ 16.2' & 5$\ $& 4$\ $\\
{ Jul. 2007}  & { 2.2m ESO/Wide Field Imager}	   & { 8 of 2k $\times$
  4k}&~~~0.238 &~34'$\times$ 33' &  7$\ $& 7$\ $\\
{ Jul. 2008}  & { 2m LT/Ratcam optical CCD camera}	   & {  2k $\times$ 2k}	       &~~~0.270$^a$&~4.6' $\times$ 4.6' & 24$^b$& 22$^b$\\
{ Jan-Jun 2009}  & { 2m Faulkes Telescope North/EM01}	   & { 2k $\times$ 2k}&~~~0.278$^a$ &~4.7'$\times$ 4.7' &  14$^c$& 7$^d$\\
\hline
	 \end{tabular}
$^{a}$ Binning 2$\times$2\\
$^{b}$ We covered 6 different fields around the galaxy center published by Belokurov et al. (2007)\\
$^{c}$ We covered 4 different fields around the galaxy center\\
$^{d}$ We covered 2 different fields around the galaxy center\\
	    \end{table}

\begin{table}
\scriptsize
\caption[]{Identification and properties of variable stars in the Hercules UFD galaxy}
\label{t:RR}

\begin{tabular}{lcclllccrr}
\hline
\hline
\noalign{\smallskip}
{\rm Name$^{a}$}&$\alpha$&$\delta$&{\rm Type}&~~{\rm P}&~~Epoch$^{b}$
&$\langle V \rangle$&$\langle B \rangle$& $A_V$~~     &$A_B$~~\\
~~        &        (2000)&(2000)&          &~  (days)&($-$2450000)&(mag)&(mag)&(mag)&(mag)\\
\noalign{\smallskip}
\hline
\noalign{\smallskip}
{\rm V1} &16:31:02.17&$+$12:47:33.7&{\rm RRab}&0.639206&4870.149&21.27&21.68&1.06&1.16 \\   
{\rm V2} &16:31:02.91&$+$12:45:48.5&{\rm  AC}    &0.53777&4614.590&20.72&21.14&0.45&0.59 \\         
{\rm V3} &16:30:54.93&$+$12:47:04.2&{\rm RRc}   &0.39997&4614.6767&21.32&21.72&0.48&0.61\\     
{\rm V4} &16:30:56.14&$+$12:48:29.2&{\rm RRc}   &0.39576&4612.5745&21.23&21.59&0.58&0.69 \\      
{\rm V5} &16:30:52.28&$+$12:49:12.0&{\rm RRc}   &0.40183&4212.699 &21.30&21.64&0.47&0.61 \\       
{\rm V6} &16:30:52.41&$+$12:49:60.0&{\rm RRab}&0.69981&4232.700 &21.35&21.76&0.90&1.19 \\    
{\rm V7} &16:31:29.48&$+$12:47:34.9&{\rm RRab}&0.67799&4613.456 &21.22&21.65&0.81&1.03\\     
{\rm V8} &16:31:27.20&$+$12:44:16.7&{\rm RRab}&0.66234&4613.497 &21.25&21.69&0.90&1.09\\	  
{\rm V9} &16:31:29.50&$+$12:40:03.1&{\rm RRab}&0.72939&4214.6101&21.18&21.60&0.80&1.00\\   
{\rm V10}$^{c}$&16:30:03.96 &$+$12:52:06.3&{\rm RRab}&0.6616 &4210.679&21.28&21.69&1.17&1.32\\ 
\hline

\end{tabular}

$^{\rm  a}$ We have named the variable stars with an increasing number  starting from \citet{Sand09} center for Hercules\\ 
$^{\rm  b}$ Epoch corresponds to the time of maximum light.\\
$^{\rm  c}$ This star has very scattered light curves\\

\end{table}

\begin{table}
\begin{center}
\caption{$B,V$ photometry of the variable stars detected in the
  Hercules UFD. This Table is published in its entirety in the electronic edition of the Journal. 
A portion is shown here
for guidance regarding its form and content.}
\label{t:pho}
\vspace{0.5 cm} 
\begin{tabular}{cccccc}
\hline
\hline
\multicolumn{6}{c}{Hercules - Star V8 - {\rm RRab}} \\
\hline
{\rm HJD}  & {\rm V} & {\rm $\sigma_V$} & {\rm HJD } & {\rm B}&{\rm $\sigma_B$} \\ 
($-$2454211) & {\rm (mag)} & {\rm (mag)}  & ($-$2454211) & {\rm (mag)} & {\rm (mag)}  \\
\hline
0.652005 &  21.26& 0.02  &  0.640218 &  21.69&      0.02 \\
0.676609 &  21.31& 0.02  &  0.664420 &  21.72&      0.02 \\
0.702719 &  21.34& 0.02  &  0.728613 &  21.89&      0.02 \\
0.741483 &  21.42& 0.02  &  1.682154 &  22.10&      0.02 \\
1.693693 &  21.59& 0.02  &  1.705106 &  22.00&      0.02 \\
1.717861 &  21.33& 0.01  &  1.730836 &  21.49&      0.01 \\
2.646191 &  21.27& 0.01  &  2.634102 &  21.69&      0.02 \\
2.686550 &  21.36& 0.02  &  2.658033 &  21.75&      0.02 \\
2.722461 &  21.42& 0.02  &  2.734216 &  21.94&      0.02 \\
3.610141 &  21.48& 0.02  &  3.621931 &  22.01&      0.02 \\

\hline 

\end{tabular}

\end{center}

\end{table}

\begin{table}
\begin{center}
\caption{Fourier parameters of the $V$-band light curve for variable stars in
 Hercules} \label{t:fou}
$$
\begin{array}{llcccccc}
\hline
\hline
{\rm Name}  &{\rm Type}&A_{21} & A_{31} &\phi_{21}& \sigma_{\phi_{21}}&\phi_{31}&
\sigma_{\phi_{31}}\\
\hline
V1 &  {\rm RRab}&0.49917 &  0.36365     &3.64 &0.10 &1.45& 0.14\\ 
V2  &  {\rm  AC   } &0.29909 &  0.1525  &4.57 &0.14 &2.0 & 0.3\\
V3 &   {\rm RRc  }&0.03663 &  0.09088   &4.4  &1.5  &3.0 & 0.6\\
V4 &   {\rm RRc  } &0.27846 &  0.11197  &4.4  &0.2  &2.3 & 0.5\\
V5 &   {\rm RRc  } &0.22652 &  0.08833  &5.10 &0.12 &3.6 & 0.3\\
V6 &   {\rm RRab}&0.54905 &  0.34341    &4.06 &0.12 &2.14& 0.19\\
V7 &   {\rm RRab}&0.39660 &  0.32364    &3.98 &0.08 &1.82& 0.10\\
V8 &  {\rm RRab}& 0.44746 &  0.34821    &3.74 &0.07 &1.60& 0.10\\
V9 &   {\rm RRab}&0.44564 &  0.39424    &3.97 &0.08 &1.74& 0.10\\
V10&  {\rm RRab}& 0.50604 &  0.37673    &4.3  &0.3  &2.8 & 0.4\\
\hline

\end{array}
$$
\end{center}

\end{table}

\begin{table}
\begin{center}
\caption{Individual metallicities$^{a}$ and  reddening values derived for the RR Lyrae stars  in
  Hercules (see text for details).}
\label{t:met}
$$
\begin{array}{lccc}
\hline
\hline
{\rm Name}  &  {\rm [Fe/H]_{ZW}} \pm \sigma_{\rm [Fe/H]_{ZW}} &
{\rm [Fe/H]_{C09}} \pm \sigma_{\rm [Fe/H]_{C09}} & E(B-V)\\
\hline
V1 &  -2.23\pm 0.13 &  -2.4  \pm 0.2 &	 0.10 \\ 
V3 & -2.12 \pm 0.06     & -2.29 \pm 0.13 & -- \\ 
V4 &  -2.13	 \pm 0.04     & -2.31 \pm 0.12 & -- \\
V5 &-2.03	 \pm 0.07     & -2.14 \pm 0.14 & -- \\
V6 &  -1.81	 \pm 0.18     & -1.8  \pm 0.3 &0.06\\
V7 & -2.02	 \pm 0.10     & -2.13 \pm 0.19 & 0.08\\
V8 & -2.18 \pm  0.10     & -2.36 \pm 0.19 & 0.12\\
V9 & -2.29	\pm 0.10     & -2.52 \pm 0.19 & 0.08\\
\hline
\hline
\end{array}
$$
\end{center}
\scriptsize
$^{\rm a}$ We could not estimate metallicity and reddening values for V10,  due to the  large scatter  of the light curves\\

\end{table}

\begin{table}
\begin{center}
\caption{Literature values for Hercules mean metallicity.} \label{t:lit_met}

\begin{tabular}{lc}
\hline
\hline
Authors  & ${\rm [Fe/H] \pm \sigma_{[Fe/H]}}$ \\
                        &  dex    \\
\hline
\citet{Col07}    &$\sim -2.26 $\\
\citet{Sand09} &$\sim -2.00 $\\
\citet{SG07}     & $-2.7\pm0.1^{a}$\\ 
\citet{Kirby08} &$-2.58 \pm 0.51 $\\
\citet{Koch08} &$\sim -2.00$\\
\citet{Aden09} &$-2.35 \pm 0.31$\\
\hline

\end{tabular}
\end{center}
\scriptsize
{$^a$} Transformed to the \citet{Carretta09} scale.

\end{table}

\end{document}